\documentclass[11pt,a4paper]{article}

\usepackage[USenglish]{babel}

\usepackage[top=2.5cm,right=2.5cm,bottom=2.5cm,left=2.5cm]{geometry}
\usepackage{setspace}
\usepackage[left]{lineno}
\usepackage[ansinew]{inputenc}
\usepackage{amsfonts}
\usepackage{amssymb,amsmath,amsthm}
\usepackage{latexsym}
\usepackage{color}

\newcommand {\debeq}	{\begin{eqnarray*}}
\newcommand {\fineq}	{\end{eqnarray*}}
\newcommand {\lbd}	{\lambda}

\def\Sp{{\mathop{\overline{X}}}}
\def\If{{\mathop{\underline{X}}}}

\def\Spy{{\mathop{\overline{Y}}}}
\def\Ify{{\mathop{\underline{Y}}}}

\newcommand	{\intgen}	
{\int_0^\infty}

\newtheorem	{thm}		{Theorem}[section]

\newtheorem	{lem} 	[thm]	{Lemma}

\newtheorem     {rem}           {Remark}
\newtheorem	{prop}	[thm]{Proposition}
\newtheorem	{cor}		[thm]{Corollary}

\newcommand	{\indic}	[1]
{{\bf{1}}_{\{#1\}}}

\newcommand{\tanjahelen}{\textcolor{black}}

\newcommand{\tanja}{\textcolor{black}}
\newcommand{\helen}{\textcolor{black}}
\newcommand{\amaury}{\textcolor{black}}
\newcommand{\amauryy}{\textcolor{black}}

\begin{document}
\linenumbers

\title{Phylogenetic analysis accounting for age-dependent death and sampling with applications to epidemics}
\author{\textsc{By Amaury Lambert$^{\ \star,1,2}$, Helen K. Alexander$^3$, and Tanja Stadler$^4$}
}
\date{}
\maketitle

\noindent
$^\star$ Corresponding author\\
\noindent
\textsc{$^1$
UPMC Univ Paris 06\\
Laboratoire de Probabilités et Modèles Aléatoires CNRS UMR 7599}\\
\noindent\textsc{$^2$
Collège de France\\
Center for Interdisciplinary Research in Biology CNRS UMR 7241\\
Paris, France}\\
\textsc{Phone number: }+33 1 44 27 13 91\\
\textsc{E-mail: }amaury.lambert@upmc.fr\\
\textsc{URL: }http://www.proba.jussieu.fr/pageperso/amaury/index.htm\\
\\
\noindent\textsc{$^3$Institute of Integrative Biology\\
ETH Z\"{u}rich\\
Universit\"{a}tsstrasse 16\\
8092 Z\"{u}rich\\
Switzerland}\\
\textsc{E-mail: }helen.alexander@env.ethz.ch\\
\textsc{URL: }http://www.tb.ethz.ch/people/helenal
\\
\\
\noindent\textsc{$^4$Institute of Integrative Biology\\
ETH Z\"{u}rich\\
Universit\"{a}tsstrasse 16\\
8092 Z\"{u}rich\\
Switzerland}\\
\textsc{E-mail: }tanja.stadler@env.ethz.ch\\
\textsc{URL: }http://www.tb.ethz.ch/people/tstadler

\begin{abstract}
\noindent
The reconstruction of phylogenetic trees based on viral genetic sequence data sequentially sampled from an epidemic provides estimates of the past transmission dynamics, by fitting epidemiological models to these trees.
To our knowledge, none of the epidemiological models currently used in phylogenetics can account for recovery rates and sampling rates dependent on the time elapsed since transmission.

Here we introduce an epidemiological model where infectives leave the epidemic, either by recovery or sampling, after some random time which may follow an arbitrary distribution.

We derive an expression for the likelihood of the phylogenetic tree of sampled infectives under our general epidemiological model. The analytic concept developed in this paper will facilitate inference of past epidemiological dynamics and provide an analytical framework for performing very efficient simulations of phylogenetic trees under our model. The main idea of our analytic study is that  the non-Markovian epidemiological model giving rise to phylogenetic trees growing vertically as time goes by, can be represented by a  Markovian ``coalescent point process'' growing horizontally by the sequential addition of pairs of coalescence and sampling times.

As examples, we discuss two special cases of our general model, namely an application to influenza and an application to HIV. Though phrased in epidemiological terms, our framework can also be used for instance to fit macroevolutionary models to  phylogenies of extant and extinct species, accounting for general species lifetime distributions.
\end{abstract}  	
\medskip
\textit{Running head.} Phylogenies with age-dependent death and sampling.
\textit{Word count.} 6000 words approximately, including supporting information (appendix).\\

\section{Introduction}

Phylogenetic trees, which are reconstructed from genetic data, describe the genealogical relationships within a population.  The analysis of these trees can provide important insights into the underlying population dynamic processes.  For instance, in a group of species descending from a common ancestor, one can construct a tree based on homologous gene(s) sequenced from these species, and thus infer speciation and extinction rates \cite{Nee1994}.  As another example, viral genetic sequences extracted from patient samples can provide information on the rate at which an infectious disease transmits in the host population \cite{Stadler2012MBE-R0}.

Maximum likelihood and Bayesian inference  are common techniques for estimating such parameters, given a model of the underlying population dynamics.  However, the complexity of models that can be applied is limited by the need to derive the likelihood of a phylogenetic tree.  Until recently, approaches were limited to death rates of individuals being independent of the age of an individual (see e.g. \cite{Nee1994,Morlon2011,Stadler2011PNAS, Stadler2012ProcRoyB} for species phylogenies and \cite{Stadler2012MBE-R0, Stadler2012PNAS} for virus phylogenies).

For phylogenetic trees in which all tips are sampled at one point in time, e.g. extant species phylogenies, Lambert \cite{L} introduced a framework to calculate the likelihood of a phylogenetic tree accounting for general lifetime distributions (see also \cite{LS}). Here we build upon this approach to additionally allow for sequential sampling.
Sequential sampling allows analysis e.g.~of virus sequence data obtained throughout the course of an epidemic. 
In the model exposition and worked examples to follow, we focus on an epidemic model in which ``births'' (branching events) represent transmission events and ``deaths'' represent events of becoming non-infectious either with or without sampling.  The model also applies to non-epidemic scenarios in which individuals are sampled at different time points, for instance when dated fossils are included in a species tree.

Allowing age-dependent death/recovery and sampling agrees with the common observation that lifetimes (time being infectious, in the epidemic model) are not generally exponential, for example the infectious period of influenza typically lasts for 5-7 days according to the Center for Disease Control (http://www.cdc.gov/flu/about/disease/spread.htm). \tanja{Extending the model to age-dependent removal will allow  quantifying the death/recovery dynamics  more accurately based on genetic sequencing data, and to test whether parameter estimates (such as the basic reproductive number $R_0$}) have been biased by the more simplistic assumption of age-independent removal rates. \tanja{Furthermore, our approach will allow rapidly simulating phylogenies under age-dependent death/recovery rates even for huge epidemic outbreaks, thus allowing  efficient investigation of the impact of age-dependent rates on the structure of the phylogenetic tree.}

The structure of the paper is as follows.  First we introduce more precisely the general model of infection and sampling. The forward-in-time (vertical) process giving rise to the phylogeny is non-Markovian due to age-dependent removal rates. We then describe the jumping chronological contour process (JCCP, or simply ``contour process'' for short), a systematic way of exploring trees. 
The contour process analysis (horizontal) makes use of a Markovian process giving rise to the phylogeny  by sequentially adding pairs of coalescence and sampling times only depending on the previous sampling time. 

We proceed to apply L\'evy process theory in order to obtain explicit expressions for the Markov process transition probabilities in terms of the so-called scale function associated with the contour process.  
This leads to the key result of the paper, an explicit formula for the likelihood of a given sampled tree as a function of the parameters of the population dynamic process (Theorem \ref{thm}).

 Two worked examples then illustrate the application of the general mathematical results: the influenza model, where the lifetime of individuals is not dependent upon whether they leave the epidemic by recovery or sampling; and the HIV model, where sampling occurs after some exponential \helen{time} 
 during the (independently distributed) infectious period.  
 
\tanja{We conclude the paper by discussing future challenges in putting the theoretical framework into a computational inference tool, such that the model can be used to analyze pathogen genetic sequence data collected during an epidemic.}

\section{Model of infection and sampling}

We model by a (possibly) non-Markovian branching process the dynamics of a population of infectives. The process is assumed to start with one infected individual and has an overall time duration of $t$. 

A ``birth'' event is interpreted as the infection of a susceptible individual, where susceptibles are supposed to be in excess, so that individuals can be assumed to give birth independently (branching property) and at constant rate, say $b$. \amaury{The branching property means in particular that the population of infectives is, on average, exponentially growing or declining (or constant).}
The new infective is assumed to be infectious immediately after infection.

The ``death'' of an individual is the removal of an individual from the infective population.
Individuals may become non-infectious because of actual death, recovery, successful treatment, or behaviour changes.
Upon removal, individuals may be sampled (type 2; i.e. included into the phylogeny) or may not be sampled (type 1).


Mathematically, we can equivalently assume that the type (1 or 2) is chosen upon infection (birth) with probabilities $c_1$ and $c_2 = 1-c_1$ respectively, independently from other individuals. Individuals of type 1 live a duration distributed as $V_1$ after which they are removed by becoming non-infectious. Individuals of type 2 live a duration distributed as $V_2$ after which they are simultaneously sampled and   removed, meaning each death of an individual of type 2 coincides with a sampling event.

These assumptions are consistent with the natural framework where, for an individual who was infected $a$ time units ago (i.e., with `age' $a$), recovery occurs with the instantaneous rate $\rho_1(a)$ and sampling occurs with the rate $\rho_2(a)$, independently. This is equivalent to saying that individuals with age $a$ leave the epidemic at rate $\rho(a):=\rho_1(a)+\rho_2(a)$ \amaury{(i.e., an individual is removed at the first point of a time-dependent Poisson process with instantaneous rate $\rho$, where time is reset at birth)}, and that upon leaving the epidemic at age $a$, they leave it by recovery with probability $\rho_1(a)/\rho(a)$ and by sampling with probability $\rho_2(a)/\rho(a)$.
This is exactly the same framework as described above, if one sets for $i=1,2$,
$$
c_i:=\int_0^\infty \rho_i(z)\,e^{-\int_0^z \rho(a)\, da}\, dz\quad\mbox{ and }\quad P(V_i\in dz) := c_i^{-1}\,\rho_i(z)\,e^{-\int_0^z \rho(a)\, da}\, dz.
$$ 
 
 

Our analyses and results apply to the general model just described, but we will later use the following two cases as examples.  In the first case (influenza model), $V_1$ and $V_2$ are identically distributed, meaning the duration of infectiousness does not depend on being sampled.  In the second case (HIV model), natural infectious lifetimes are distributed as some random variable $V$, while sampling is assumed to occur after some independent exponential duration with parameter $\mu$, meaning individuals are sampled with a constant rate $\mu$ while being infectious. The type of an individual is determined by the first event to occur (removal with or without sampling). 

\begin{figure}[ht]
\unitlength 2mm 
\linethickness{0.2pt}
\ifx\plotpoint\undefined\newsavebox{\plotpoint}\fi 
\begin{picture}(73,28)(0,0)
\put(9,15){\line(0,1){6}}
\put(13,19){\line(0,1){8}}
\put(17,12){\line(0,1){5}}
\put(21,15){\line(0,1){5}}
\put(25,17){\line(0,1){11}}
\put(29,10){\line(0,1){8}}
\put(33,16){\line(0,1){9}}
\put(41,14){\line(0,1){5}}
\put(45,16){\line(0,1){4}}
\put(17,17){\circle*{1.061}}
\put(9,21){\circle*{1.061}}
\put(54,21){\circle*{1.061}}
\put(62,17){\circle*{1.061}}
\put(70,20){\circle*{1.061}}
\put(37,20){\circle*{1.061}}
\put(8.982,14.982){\line(-1,0){.8}}
\put(7.382,14.982){\line(-1,0){.8}}
\put(5.782,14.982){\line(-1,0){.8}}
\put(12.982,18.982){\line(-1,0){.8}}
\put(11.382,18.982){\line(-1,0){.8}}
\put(9.782,18.982){\line(-1,0){.8}}
\put(16.982,11.982){\line(-1,0){.9231}}
\put(15.136,11.982){\line(-1,0){.9231}}
\put(13.29,11.982){\line(-1,0){.9231}}
\put(11.444,11.982){\line(-1,0){.9231}}
\put(9.598,11.982){\line(-1,0){.9231}}
\put(7.752,11.982){\line(-1,0){.9231}}
\put(5.906,11.982){\line(-1,0){.9231}}
\put(20.982,14.982){\line(-1,0){.8}}
\put(19.382,14.982){\line(-1,0){.8}}
\put(17.782,14.982){\line(-1,0){.8}}
\put(24.982,16.982){\line(-1,0){.8}}
\put(23.382,16.982){\line(-1,0){.8}}
\put(21.782,16.982){\line(-1,0){.8}}
\put(28.982,9.982){\line(-1,0){.96}}
\put(27.062,9.982){\line(-1,0){.96}}
\put(25.142,9.982){\line(-1,0){.96}}
\put(23.222,9.982){\line(-1,0){.96}}
\put(21.302,9.982){\line(-1,0){.96}}
\put(19.382,9.982){\line(-1,0){.96}}
\put(17.462,9.982){\line(-1,0){.96}}
\put(15.542,9.982){\line(-1,0){.96}}
\put(13.622,9.982){\line(-1,0){.96}}
\put(11.702,9.982){\line(-1,0){.96}}
\put(9.782,9.982){\line(-1,0){.96}}
\put(7.862,9.982){\line(-1,0){.96}}
\put(5.942,9.982){\line(-1,0){.96}}
\put(32.982,15.982){\line(-1,0){.8}}
\put(31.382,15.982){\line(-1,0){.8}}
\put(29.782,15.982){\line(-1,0){.8}}
\put(36.982,11.982){\line(-1,0){.8889}}
\put(35.205,11.982){\line(-1,0){.8889}}
\put(33.427,11.982){\line(-1,0){.8889}}
\put(31.649,11.982){\line(-1,0){.8889}}
\put(29.871,11.982){\line(-1,0){.8889}}
\put(40.982,13.982){\line(-1,0){.8}}
\put(39.382,13.982){\line(-1,0){.8}}
\put(37.782,13.982){\line(-1,0){.8}}
\put(44.982,15.982){\line(-1,0){.8}}
\put(43.382,15.982){\line(-1,0){.8}}
\put(41.782,15.982){\line(-1,0){.8}}
\put(3,23.375){\makebox(0,0)[cc]{\scriptsize $t$}}
\put(2.875,3.375){\makebox(0,0)[cc]{\scriptsize $0$}}
\put(51.875,3.375){\makebox(0,0)[cc]{\scriptsize $0$}}
\put(7,21){\makebox(0,0)[cc]{\scriptsize $1$}}
\put(52.5,21){\makebox(0,0)[cc]{\scriptsize $1$}}
\put(15,17){\makebox(0,0)[cc]{\scriptsize $2$}}
\put(60.5,17){\makebox(0,0)[cc]{\scriptsize $2$}}
\put(36,21){\makebox(0,0)[cc]{\scriptsize $3$}}
\put(68.625,21.125){\makebox(0,0)[cc]{\scriptsize $3$}}
\put(37,12){\line(0,1){8}}
\put(5,18){\line(0,-1){15}}
\put(10,3){\vector(0,-1){.035}}\put(10,21){\vector(0,1){.035}}\put(10,21){\line(0,-1){18}}
\put(55.5,3){\vector(0,-1){.035}}\put(55.5,21){\vector(0,1){.035}}\put(55.5,21){\line(0,-1){18}}
\put(18,3){\vector(0,-1){.035}}\put(18,17){\vector(0,1){.035}}\put(18,17){\line(0,-1){14}}
\put(63.5,3){\vector(0,-1){.035}}\put(63.5,17){\vector(0,1){.035}}\put(63.5,17){\line(0,-1){14}}
\put(38,3){\vector(0,-1){.07}}
\put(71.5,3){\vector(0,-1){.07}}
\put(38,20){\vector(0,1){.07}}
\put(71.5,20){\vector(0,1){.07}}
\put(38,20){\line(0,-1){17}}
\put(71.5,20){\line(0,-1){17}}
\put(11,15){\makebox(0,0)[cc]{\scriptsize $S_1$}}
\put(56.5,15){\makebox(0,0)[cc]{\scriptsize $S_1$}}
\put(19.125,12){\makebox(0,0)[cc]{\scriptsize $S_2$}}
\put(64.625,12){\makebox(0,0)[cc]{\scriptsize $S_2$}}
\put(39,11){\makebox(0,0)[cc]{\scriptsize $S_3$}}
\put(72.5,11){\makebox(0,0)[cc]{\scriptsize $S_3$}}
\put(12,3){\vector(0,-1){.035}}\put(12,12){\vector(0,1){.035}}\put(12,12){\line(0,-1){9}}
\put(59,3){\vector(0,-1){.035}}\put(59,12){\vector(0,1){.035}}\put(59,12){\line(0,-1){9}}
\put(20,3){\vector(0,-1){.035}}\put(20,10){\vector(0,1){.035}}\put(20,10){\line(0,-1){7}}
\put(67,3){\vector(0,-1){.035}}\put(67,10){\vector(0,1){.035}}\put(67,10){\line(0,-1){7}}
\put(13.375,7.375){\makebox(0,0)[cc]{\scriptsize $R_2$}}
\put(60.375,7.375){\makebox(0,0)[cc]{\scriptsize $R_2$}}
\put(21.125,6.75){\makebox(0,0)[cc]{\scriptsize $R_3$}}
\put(68.125,6.75){\makebox(0,0)[cc]{\scriptsize $R_3$}}
\put(4,3){\line(1,0){42}}
\multiput(3.982,22.982)(.975,0){41}{{\rule{.2pt}{.2pt}}}
\put(54,3){\line(0,1){18}}
\put(62,12){\line(0,1){5}}
\put(70,10){\line(0,1){10}}
\put(19.125,26){\makebox(0,0)[cc]{a)}}
\put(62,24){\makebox(0,0)[cc]{b)}}
\put(61.982,11.982){\line(-1,0){.8889}}
\put(60.205,11.982){\line(-1,0){.8889}}
\put(58.427,11.982){\line(-1,0){.8889}}
\put(56.649,11.982){\line(-1,0){.8889}}
\put(54.871,11.982){\line(-1,0){.8889}}
\put(69.982,9.982){\line(-1,0){.9412}}
\put(68.1,9.982){\line(-1,0){.9412}}
\put(66.218,9.982){\line(-1,0){.9412}}
\put(64.335,9.982){\line(-1,0){.9412}}
\put(62.453,9.982){\line(-1,0){.9412}}
\put(60.571,9.982){\line(-1,0){.9412}}
\put(58.688,9.982){\line(-1,0){.9412}}
\put(56.806,9.982){\line(-1,0){.9412}}
\put(54.924,9.982){\line(-1,0){.9412}}
\put(53,3){\line(1,0){20}}
\end{picture}

\caption{ a) The oriented phylogeny of the epidemics showing transmission events (horizontal dashed lines) and sampling events (black dots), for 3 infectives sampled before present time $t$ (dotted line), and 3 infectives alive at time $t$; b) Sampling times ($S_i$) and coalescence times $(R_i)$ characterizing the oriented sampled tree (see main text). 
}
\label{fig:markedtree}
\end{figure}
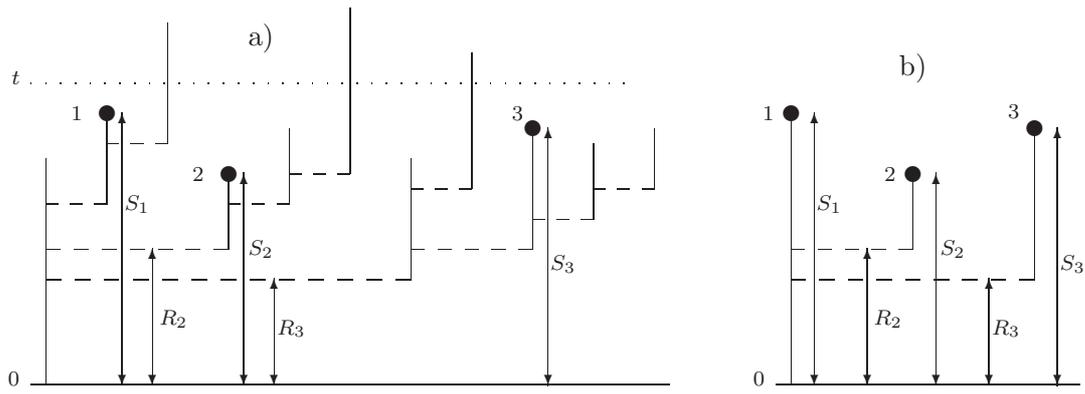

The binary random tree, embedded in continuous time, of this two-type population can be viewed as a two-type \emph{splitting tree}, where in addition the tip of every edge corresponding to the life of an individual of type 2 is \emph{marked} as a sampling point, see Figure \ref{fig:markedtree}. Splitting trees \cite{GK, L2, L, L3} are those random trees generated by a so-called \emph{homogeneous, binary Crump--Mode--Jagers process} (CMJ), that is, a branching process where individuals give birth singly and at constant rate $b$, during lifetimes that are independent and identically distributed (iid), distributed as some random variable $V$, which is not necessarily exponentially distributed. In particular, the process counting the total population size is not necessarily Markovian. The law of a splitting tree is characterized by the measure $\pi(\cdot) := bP(V\in \cdot)$ usually called the lifespan measure. 

Here, the law of our two-type splitting tree is characterized by the knowledge of the two lifespan measures $\pi_1:=bc_1P(V_1\in\cdot)$ and $\pi_2:=b(1-c_1)P(V_2\in\cdot)$. Notice that regardless of types/marks, the genealogical tree of the whole population is a splitting tree with lifespan measure $\pi:=\pi_1+\pi_2$.




We call the \emph{sampled tree} the part of the marked splitting tree which is \emph{spanned by its marks and the root}, that is, the phylogenetic tree of samples (i.e., when all lineages without sampled descendants are pruned). \amaury{See Figure \ref{fig:markedtree}b for a graphical representation.} Assuming that the sampled tree can be reconstructed exactly from the patient samples, our goal is to provide a method for computing \helen{the probability density (likelihood) of a  sampled tree for given parameters under our model.  The method can also be used to compute the posterior likelihood of the parameters given the data, in a Bayesian framework where parameters are given a prior distribution.
The likelihood allows us to infer parameters of the epidemiological process from the sampled tree using maximum likelihood or Bayesian methodology.} 

From now on, we assume that the tree is embedded in the plane, employing the natural \emph{orientation} where each daughter edge sprouts to the right of its mother edge (see Figure \ref{fig:markedtree}). Our next step is to describe a process which allows us to systematically explore plane splitting trees, and elucidates how plane sampled trees under our model may be represented simply by pairs of coalescence and sampling times. 

\section{The Contour Process}

\begin{figure}[ht]
\unitlength 2mm 
\linethickness{0.2pt}
\ifx\plotpoint\undefined\newsavebox{\plotpoint}\fi 
\begin{picture}(79,26)(0,0)
\put(6,21){\circle*{1.061}}
\put(19,17){\circle*{1.061}}
\put(50,20){\circle*{1.061}}
\put(3,18){\line(1,-1){3}}
\put(6,15){\line(0,1){6}}
\put(6,21){\line(1,-1){2}}
\put(8,19){\line(0,1){4}}
\put(8,23){\line(1,-1){11}}
\put(19,12){\line(0,1){5}}
\put(19,17){\line(1,-1){2}}
\put(21,15){\line(0,1){5}}
\put(21,20){\line(1,-1){3}}
\put(24,17){\line(0,1){6}}
\put(24,23){\line(1,-1){13}}
\put(37,10){\line(0,1){8}}
\put(37,18){\line(1,-1){2}}
\put(39,16){\line(0,1){7}}
\put(39,23){\line(1,-1){11}}
\put(50,12){\line(0,1){8}}
\put(50,20){\line(1,-1){6}}
\put(56,14){\line(0,1){5}}
\put(56,19){\line(1,-1){3}}
\put(59,16){\line(0,1){4}}
\put(59,20){\line(1,-1){17}}
\put(3,3){\vector(0,1){23}}
\multiput(2.982,22.982)(.984127,0){64}{{\rule{.2pt}{.2pt}}}
\put(1.625,23.25){\makebox(0,0)[cc]{\scriptsize $t$}}
\put(2,3){\vector(1,0){77}}
\put(6,2.5){\line(0,1){1}}
\put(8,2.5){\line(0,1){1}}
\put(19,2.5){\line(0,1){1}}
\put(24,2.5){\line(0,1){1}}
\put(39,2.5){\line(0,1){1}}
\put(50,2.5){\line(0,1){1}}
\put(76,2.5){\line(0,1){1}}
\put(3,1){\makebox(0,0)[cc]{\scriptsize $0$}}
\put(6,1){\makebox(0,0)[cc]{\scriptsize $u_1$}}
\put(8,1){\makebox(0,0)[cc]{\scriptsize $u_2$}}
\put(19,1){\makebox(0,0)[cc]{\scriptsize $u_3$}}
\put(24,1){\makebox(0,0)[cc]{\scriptsize $u_4$}}
\put(39,1){\makebox(0,0)[cc]{\scriptsize $u_5$}}
\put(50,1){\makebox(0,0)[cc]{\scriptsize $u_6$}}
\put(76,1){\makebox(0,0)[cc]{\scriptsize $u_7$}}
\put(80,3){\makebox(0,0)[cc]{\scriptsize $u$}}
\put(5.982,2.982){\line(0,1){.9231}}
\put(5.982,4.829){\line(0,1){.9231}}
\put(5.982,6.675){\line(0,1){.9231}}
\put(5.982,8.521){\line(0,1){.9231}}
\put(5.982,10.367){\line(0,1){.9231}}
\put(5.982,12.213){\line(0,1){.9231}}
\put(5.982,14.059){\line(0,1){.9231}}
\put(7.982,2.982){\line(0,1){.9412}}
\put(7.982,4.865){\line(0,1){.9412}}
\put(7.982,6.747){\line(0,1){.9412}}
\put(7.982,8.629){\line(0,1){.9412}}
\put(7.982,10.512){\line(0,1){.9412}}
\put(7.982,12.394){\line(0,1){.9412}}
\put(7.982,14.277){\line(0,1){.9412}}
\put(7.982,16.159){\line(0,1){.9412}}
\put(7.982,18.041){\line(0,1){.9412}}
\put(18.982,2.982){\line(0,1){.9}}
\put(18.982,4.782){\line(0,1){.9}}
\put(18.982,6.582){\line(0,1){.9}}
\put(18.982,8.382){\line(0,1){.9}}
\put(18.982,10.182){\line(0,1){.9}}
\put(23.982,2.982){\line(0,1){.9333}}
\put(23.982,4.849){\line(0,1){.9333}}
\put(23.982,6.716){\line(0,1){.9333}}
\put(23.982,8.582){\line(0,1){.9333}}
\put(23.982,10.449){\line(0,1){.9333}}
\put(23.982,12.316){\line(0,1){.9333}}
\put(23.982,14.182){\line(0,1){.9333}}
\put(23.982,16.049){\line(0,1){.9333}}
\put(38.982,2.982){\line(0,1){.9286}}
\put(38.982,4.84){\line(0,1){.9286}}
\put(38.982,6.697){\line(0,1){.9286}}
\put(38.982,8.554){\line(0,1){.9286}}
\put(38.982,10.411){\line(0,1){.9286}}
\put(38.982,12.268){\line(0,1){.9286}}
\put(38.982,14.125){\line(0,1){.9286}}
\put(49.982,2.982){\line(0,1){.9091}}
\put(49.982,4.801){\line(0,1){.9091}}
\put(49.982,6.619){\line(0,1){.9091}}
\put(49.982,8.437){\line(0,1){.9091}}
\put(49.982,10.255){\line(0,1){.9091}}
\put(49.982,12.073){\line(0,1){.9091}}
\end{picture}

\caption{
The marked contour process, with jumps in solid line, which is associated to the marked tree of Figure \ref{fig:markedtree}. Exploration time is denoted by $u$, and times $u_1$ to $u_6$ are all jump times of the contour process corresponding to lifetimes of individuals who are either alive at $t$ or sampled before $t$. The process terminates at time $u_7$.
 }
\label{fig:markedjccp}
\end{figure}
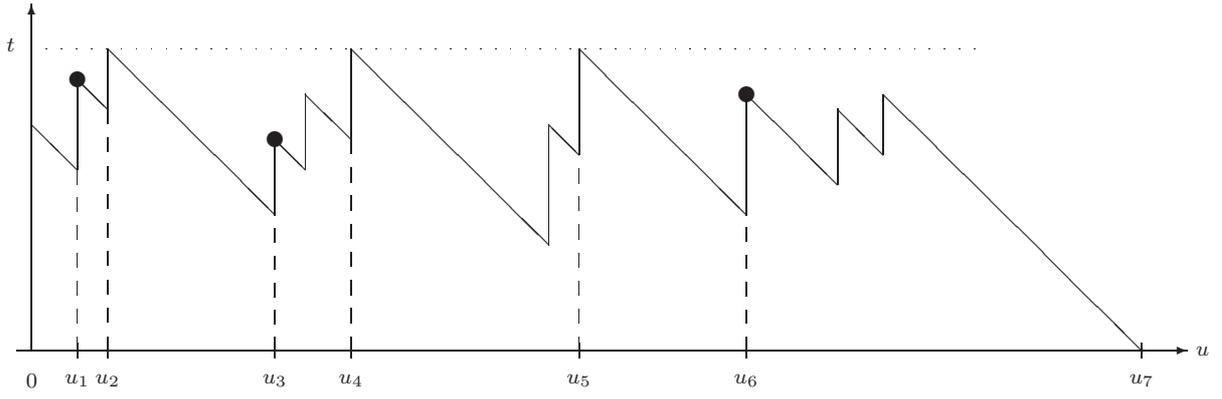

\tanjahelen{
In \cite{L}, Lambert has considered the so-called jumping chronological contour process (JCCP), or simply contour process, of the plane splitting tree truncated up to height (time) $t$. \amaury{This process can be seen as the path of a ball that follows an outline of the oriented tree, decreasing at unit speed along 
its edges \amauryy{(which are vertical and embedded in the plane)}, and jumping instantaneously to the tip of the daughter edge when reaching a node.
Figure \ref{fig:markedjccp} shows the contour process associated to the tree in Figure \ref{fig:markedtree}a.}
}


The contour process can also be seen as an alternative representation of the transmission process. The ball starts at the ``death'' of an infective and slips back until the corresponding infective transmits. Due to transmission being a Poisson process, we can have the ball slip backward in time until transmission, rather than forward in time until transmission. At transmission, the ball jumps to the time of ``death'' of the newly infected individual, and again the ball slips back until the next transmission occurs. Once the ball reaches the time of infection of the current infective, it returns to the donor in the infection event of consideration.

Observe that the number of visits of $t$ by the contour process is exactly the number of individuals in the population at time $t$. Details can be found in \cite{L,L3}. We now seek to uncover the law of this process under our model.


Now let $X$ denote the stochastic process with derivative $-1$ almost everywhere, which jumps at rate $bc_1$, with jump sizes distributed as $V_1$. In probabilistic terms, $X$ is a \emph{compound Poisson process} with \emph{jump measure} $\pi_1$ \emph{compensated} at rate $-1$. In the absence of sampled individuals, 
we have shown \cite[Theorem 4.3]{L} that the contour process has exactly the same law as the process $X$ reflected below $t$ (meaning sent back to exactly $t$ whenever it overshoots), and killed upon hitting 0. 

From now on, $X$ will denote this stochastic process, which properly reflected and killed, is the contour process of the population \emph{on unsampled individuals}.  \amauryy{The idea is that the subpaths between sampled individuals, into which we will later break up the process, can be seen as independent realizations of $X$.} We denote the law of $X$ by $P$, writing $P_x$ when conditioning on $X_0=x$. Nevertheless, unless otherwise specified, the denomination `contour process' will be reserved for the contour process of the \emph{whole} population.

Now when we additionally consider sampled individuals, 
recall that regardless of their types, individuals give birth to type 2 individuals at rate $bc_2$. Since the contour process visits the tree at unit speed, by the lack-of-memory property of the exponential distribution, it is easy to see that the contour process of the two-type splitting tree can be obtained from $X$ by \emph{adding} jumps, whose sizes are distributed as $V_2$, and which occur after independent exponential random variables with parameter $bc_2$ (further reflecting this new process under $t$ and killing it upon hitting 0). By analogy with the representation in Figure \ref{fig:markedjccp}, we will call these jumps the \emph{marked jumps} of the contour process. It is straightforward that this new process is just the compound Poisson process with jump measure $\pi$ compensated at rate $-1$.  However, we stick to the previous two-type description in order to keep track of births of type 2 individuals.

\section{The 2D coalescent point process}
In this section we show that pairs of consecutive sampling times and coalescence times in the sampled phylogeny extracted from the contour process give rise to a so-called \emph{coalescent point process}. This observation will allow us to provide an expression for the probability of the sampled tree.

Assume that we label sampled individuals (i.e., type 2 individuals) $1,2,\ldots$ in the order of the contour, that is, from left to right. We denote by $S_i$ the \emph{sampling time} of individual $i$, which is, by assumption, the (only) time at which this individual is sampled. We further denote by $R_i$ the \emph{coalescence time} between individuals $i-1$ and $i$, that is, the time at which their most recent common ancestor in the epidemic transmitted the disease to an  ancestor of $i$ (which can be assumed, for practical applications, to also be the coalescence time between the pathogens carried by $i-1$ and $i$). 

Our first remark is that the pairs $(R_i,S_i)$ characterize the (plane) sampled tree, as seen in Figure \ref{fig:markedtree}. By analogy with  phylogenies spanned by extant individuals (where one can consider $S_i = t$ for all $i$), we will say that $(R_i,S_i)$ form a \emph{two-dimensional coalescent point process} \cite{AP, L, LS}.  
Straightforward consequences of the definition of the JCCP are the following:
\begin{enumerate}
 \item the sampling time $S_i$ is the value of the contour process at its $i$-th marked jump;

 \item the coalescence time $R_i$ is the infimum of the contour process between the ($i-1$)-th and the $i$-th marked jump. 
\end{enumerate} 
In the  special case when the progenitor is sampled (before time $t$), $S_1$ is actually the lifetime of the progenitor (which can be seen as the jump size of a marked jump at exploration time $0$).
  
Now by the Markov property of the contour process, the pairs $(R_i, S_i)$ form a \emph{killed Markov chain}, where the transition probability only depends on the second component, \amaury{thanks to the fact that $S_i$ is the new starting point of the marked contour pocess. } \amaury{A killed Markov chain is a Markov chain with a possibly finite (random) lifetime. More specifically, the transition kernel $p(x,\cdot)$ of a killed Markov chain $X$ with values in some space $E$ is a sub-probability kernel, in the sense that $p(x,E)\le 1$. Then at each time step $n$, conditional on $X_n=x$, the Markov chain is killed (has lifetime $n$) with probability $1-p(x,E)$, and with probability $p(x,E)$, makes a transition according to the probability kernel $p(x,\cdot)/p(x,E)$.}

We now characterize the transitions of this Markov chain in terms of the contour process.
To get rid of the reflection at $t$, we apportion the path of the contour process into all subpaths terminating as soon as a marked jump appears or as the path exits \tanjahelen{$(0,t]$}. We classify subpaths according to the four following events: 
\begin{itemize}
\item[$A$] --
exit from the bottom of \tanjahelen{$(0,t]$}, i.e.~hit 0, before the first marked jump; 
\item[$B$] -- 
arrival of a marked jump with terminal value in \tanjahelen{$(0,t]$} before exit of \tanjahelen{$(0,t]$};
\item[$C$] --   exit from the top of \tanjahelen{$(0,t]$}, i.e.~overshoot $t$, strictly before the \amaury{next} marked jump;
\item[$C'$] -- exit from the top of \tanjahelen{$(0,t]$} \emph{at} the \amaury{next} marked jump. 
\end{itemize}
Notice that the events $A, B, C, C'$ form a partition.
The path on Figure \ref{fig:markedjccp} is apportioned into 7 subpaths delineated by times $0<u_1<\cdots <u_7$, which are respectively of types $B$, $CC'$, $B$, $CC'$, $CC'$, $B$, $A$ (where $CC'$ means: $C\cup C'$). Generally speaking, the last subpath, and the last subpath only, is of type $A$, and by the Markov property, there is a geometric number of subpaths of type $CC'$ between two consecutive marked jumps, each corresponding to subpaths of type $B$. Our objective is now to compute the joint law of the infimum of this concatenation of subpaths of type $CC'$ and of the terminal value of the concluding subpath of type $B$.

Recall that the first marked jump occurs after an exponential random variable that we denote by $\mathbf{e}$, with rate parameter 
$$q:=bc_2,$$
at which time the contour process has a jump distributed as $V_2$. \amaury{Now recall that $X$ denotes the contour process in the absence of sampled individuals}.  \helen{Throughout the paper, $V_2$ is assumed independent of $\mathbf{e}$ and $X$.}
We denote by $T_{\cal A}$ the first hitting time of the set ${\cal A}$ by $X$, and by $T := T_0\wedge T_{(t,+\infty)}$ the first exit time of $(0,t]$ by $X$, \amaury{where we use the notation $a\wedge b=\min (a,b)$}. We further denote by $\If$ and $\Sp$, respectively, the infimum and supremum processes of $X$, that is,
$$
\If_s = \inf_{0\le u\le s} X_u\quad \mbox{ and }\quad \Sp_s = \sup_{0\le u\le s}X_u. 
$$
We now express the events $A$, $B$, $C$, $C'$ using the preceding notation. \amaury{By the strong Markov property of $X$, it is sufficient to characterize each of these events in terms of one single path of $X$:} 
$$
A=\{T_0<T_{(t,+\infty)}\wedge \mathbf{e}\},
$$
$$
B= \{\mathbf{e}< T,X_{\mathbf{e}}+V_2 \tanjahelen{\leq} t\},
$$
$$
C=\{T_{(t,+\infty)}<T_0\wedge \mathbf{e}\},
$$
$$
 C'= \{\mathbf{e}<T, X_{\mathbf{e}}+V_2>t\} = \{\If_{\mathbf{e}}>0, \Sp_{\mathbf{e}}\le t, X_{\mathbf{e}}+V_2>t\}.
$$
On $C\cup C'$, it will sometimes be useful to call $\tau$ the first time at which the contour process is reflected.
 
Recall that the pairs $(R_i, S_i)$ form a Markov chain, where the transition probability only depends on the second component, which also is the new starting point of the contour. For this reason, we will define the pair $(R,S)$ by 
$$
P_x(R\in dy, S\in dz):=P(R_2\in dy, S_2\in dz\,|\, S_1=x)=:p(x; dy\, dz).
$$
At each  step $i$, conditional on $S_i=x$, the Markov chain can be killed with probability
$$
k(x)=1-\int_{[0,x]}\int_{[y,t]} p(x; dy\, dz).
$$
\amaury{Now we decompose the contour process into its excursions} \helen{(subpaths) below $t$ until the first excursion, say $\epsilon$, hitting 0 (killing, 
type $A$)} \amaury{or possessing a marked jump (sampling, type $B$). In particular, 
\begin{itemize}
\item
$S\in dz$ if the first excursion
 of type $A\cup B$ is actually of type $B$ and its marked jump ends in $dz$;
\item
$R>y$ if the infimum of the contour process until $\epsilon$ is larger than $y$, where the infimum has to be taken over the geometrically distributed number of excursions of the process (type $C\cup C'$) preceding $\epsilon$;
\item
$\epsilon$ can either be the very first excursion (starting from $x$) or any other excursion (starting from $t$). 
\end{itemize}
Therefore, we have
\begin{multline*}
P_x(R>y, S\in dz) = P_x (\If_{\mathbf{e}}> y, \Sp_{\mathbf{e}}\le t, X_{\mathbf{e}} +V_2\in dz)\\
+P_x(\If_\tau > y, C\cup C')\,\sum_{n\ge 0}\big(P_t(\If_\tau > y, C\cup C')\big)^{n}\,P_t (\If_{\mathbf{e}}> y, \Sp_{\mathbf{e}}\le t, X_{\mathbf{e}} +V_2\in dz) .
\end{multline*}
With the same line of reasoning,
$$
k(x) = P_x(A)+P_x(C\cup C')\, \sum_{n\ge 0} \big(P_t ( C\cup C')\big)^n \, P_t(A)
$$}
\helen{Rewriting the summations, we arrive at the following statement.}
\begin{prop}
\label{prop : prelim}
Let $x\in(0,t]$, $y\in [0,x)$ and $z\in (y,t)$. Then
\begin{multline*}
P_x(R>y, S\in dz) = P_x (\If_{\mathbf{e}}> y, \Sp_{\mathbf{e}}\le t, X_{\mathbf{e}} +V_2\in dz)\\
+P_x(\If_\tau > y, C\cup C')\,\frac{P_t (\If_{\mathbf{e}}> y, \Sp_{\mathbf{e}}\le t, X_{\mathbf{e}} +V_2\in dz)}{1-P_t(\If_\tau > y, C\cup C')} ,
\end{multline*}
and
$$
k(x)= P_x(A)+P_x(C\cup C')\, P_t(A\,|\,A\cup B) .
$$
\end{prop}

We will use the fact that $X$ is a L\'evy process in order to obtain explicit expressions for the above probabilities, finally leading to an explicit expression for the probability of a sampled tree in Theorem \ref{thm}.  In the following section, we first introduce the necessary background results on L\'evy processes.

\section{Lévy processes and scale functions} \label{SectionLevy}



The standard results presented in this section can be found in \cite{B, B2, L3}.  We state these results in terms of an arbitrary compound Poisson process $Y$ with jump measure $\pi$ on $(0,+\infty)$ with total mass $b$, compensated at rate $-1$. We stick to the notation defined earlier for $X$ (law $P_x$ when started from $x$, first hitting time $T_{\cal A}$ of $\cal A$ and extremum processes $\Ify$ and $\Spy$). It can be convenient to characterize the law of this process  by its Laplace exponent $\psi$ defined by
\begin{equation}
\label{eqn : psi}
\psi(\lbd):= \lbd -\intgen \pi(dx) (1-e^{-\lbd x}) \qquad \lbd\ge 0.
\end{equation}
The function $\psi$ is differentiable and convex and we denote by $\eta$ its largest root. Then $\psi$ is increasing on $[\eta,+\infty)$  and we denote by $\phi$ its inverse on this set, so that $\phi$ is a bijection from $[0,\infty)$ to $[\eta,\infty)$.


The probability of exit of an interval (from the bottom or from the top) by $Y$  has a simple expression (see e.g.~\cite{B}), in the form
\begin{equation}
\label{eqn : two-sided}
P_x(T_0< T_{(t,+\infty)}) = \frac{W(t-x)}{W(t)},
\end{equation}
where the so-called \emph{scale function} $W$ is the non-negative, nondecreasing, differentiable function such that $W(0)=1$, characterized by its Laplace transform 
\begin{equation}
\label{eqn : LT scale}
\intgen dx\, e^{-\lbd x} \, W(x) = \frac{1}{\psi(\lbd)} \qquad \lbd>\eta.
\end{equation}
Equation \eqref{eqn : two-sided} gives the probability that $Y$ exits $(0,t]$ from the bottom of the interval. The following formula gives the Laplace transform of $T$ on this event.

 For any $q>0$,
\begin{equation}
\label{eqn : q-two-sided}
E_x\left(e^{-qT} \,\mathbf{1}_{\{T_0< T_{(t,+\infty)}\}}\right) = \frac{W^{(q)}(t-x)}{W^{(q)}(t)},
\end{equation}
where the so-called $q$-\emph{scale function} $W^{(q)}$ is the non-negative, nondecreasing, differentiable function such that $W^{(q)}(0)=1$, characterized by its Laplace transform 
\begin{equation}
\label{eqn : LT q-scale}
\intgen dx\, e^{-\lbd x} \, W^{(q)}(x) = \frac{1}{\psi(\lambda)-q} \qquad \lambda>\phi(q).
\end{equation}
Note that $W^{(0)} \equiv W$. Last, the $q$-resolvent of the process killed upon exiting \tanjahelen{$(0,t]$} is given by the following formula
\begin{equation}
\label{eqn : q-resolvent}
u^q_t(x,z)\,dz:=E_x\left(\int_{s=0}^T ds\,e^{-qs}\mathbf{1}_{\{Y_s\in dz\}}\right) = \frac{W^{(q)}(t-x)\,W^{(q)}(z)}{W^{(q)}(t)} - \indic{z\ge x}W^{(q)}(z-x) .
\end{equation}
Observe that by the Fubini--Tonelli Theorem
\begin{equation}
\label{eqn : q-resolvent-bis}
qu^q_t(x,z)\,dz= E_x\left(\int_{s=0}^T \mathbf{1}_{\{\mathbf{e}\in ds\}}\,\mathbf{1}_{\{Y_s\in dz\}}\right)
=
P_x\left(\mathbf{e}<T, Y_\mathbf{e}\in dz\right)
=P_x (\Ify_{\mathbf{e}} >0, \Spy_{\mathbf{e}} \tanjahelen{\leq} t, Y_{\mathbf{e}}\in dz),
\end{equation}
where $\mathbf{e}$ denotes an independent exponential random variable with parameter $q$.
The previous formula is key to computing the probabilities involved in Proposition \ref{prop : prelim} (see Appendix).
%
%
%
We will use the following useful lemma (proved in the Appendix) several times.
\begin{lem}
\label{lem:useful} For any $z,q\ge 0$, 
$$
\int_0^z W^{(q)}(z-x) \,\pi(dx) = (q+b)W^{(q)} (z) - W^{(q)\prime}(z).
$$
\end{lem}


\section{The likelihood of the sampled tree }

We now apply the  results from Section \ref{SectionLevy} to the process $X$ (the contour process on nonsampled individuals), in order to give an explicit formula for the probabilites displayed in Proposition \ref{prop : prelim}.
Let $\psi_1$ be the Laplace exponent of $X$:
$$
\psi_1(\lambda) = \lbd -\intgen bc_1P(V_1\in dx) (1-e^{-\lbd x}),
$$
and $W_1^{(q)}$ the $q$-scale function associated with $\psi_1$ and defined in \eqref{eqn : LT q-scale}, required now for the specific $q = b c_2$. Note that all formulae given in the previous section hold for a general $q$, but that from now on we will always assume $q=b c_2$.
We will use the following definitions
\begin{equation}
\label{eqn : dfn C}
C_1^{(q)}(z) := q\int_0^z W_1^{(q)}(z-u) P(V_2\in du),
\end{equation}
and
\begin{equation}
\label{eqn : dfn U}
U_1^{(q)}(z):=1+\int_0^z C_1^{(q)}(x)\, dx= 1+q\int_0^z W_1^{(q)}(z-u) P(V_2\le u)\, du.
\end{equation}
\helen{The last equality comes from an application of Fubini--Tonelli theorem and a change of variable.  Notice in particular that $U_1^{(q)\prime} = C_1^{(q)}$.}
Then we have the following results, for which proofs can be found in the Appendix.
\begin{lem}
\label{lem1}
Let $x\in(0,t]$, $y\in [0,x)$ and $z\in (y,t)$.
Then 
$$
P_x(\If_{\mathbf{e}}> y, \Sp_{\mathbf{e}}\le t, X_{\mathbf{e}} +V_2\in dz)
=
\left(\frac{W_1^{(q)}(t-x)}{W_1^{(q)}(t-y)}C_1^{(q)}(z-y)-\indic{z\ge x}C_1^{(q)}(z-x)\right)\, dz.
$$
\end{lem}

\begin{lem}
\label{lem2}
Let $x\in(0,t]$ and $y\in [0,x)$.
Then 
$$
P_x(\If_\tau > y, C\cup C')
= U_1^{(q)}(t-x)  - \frac{W_1^{(q)}(t-x)}{W_1^{(q)}(t-y)}U_1^{(q)}(t-y).
$$
\end{lem}

We can now state the main result of this article.
\begin{thm}
\label{thm}
The sequence $S_1, (R_2, S_2), (R_3, S_3),\ldots$ is a killed Markov chain where the transition probability only depends on  the second component ($S_i$), and for any $x\in(0,t]$, $y\in [0,x)$ and $z\in (y,t)$, the starting point has distribution
$$
P(S_1\in dz) 
= c_2P(V_2\in dz) + \left(c_2 \int_0^z  P(V_2\in du)\, W_1^{(q)\prime}(z-u)-\frac{C_1^{(q)}(z)C_1^{(q)}(t)}{bU_1^{(q)}(t)}\right)\, dz,
$$
the transition probability $p(x; dy \, dz)= P_x( R\in dy, S \in dz)$ is characterized by
$$
P_x(R> y, S \in dz) = \left(C_1^{(q)}(z-y)\,\frac{U_1^{(q)}(t-x)}{U_1^{(q)}(t-y)}-\indic{z\ge x}C_1^{(q)}(z-x)\right)\, dz ,
$$
and the killing probability is 
\begin{equation}
\label{k}
k(x)=\frac{U_1^{(q)}(t-x)}{U_1^{(q)}(t)} .
\end{equation}
The probability $p$ that at least one individual is sampled before time $t$ (i.e., the sequence is not empty), is given by 
\begin{equation}
\label{p}
p= \int_0^t P(S_1\in dz) =
\frac{C_1^{(q)}(t)}{bU_1^{(q)}(t)}. 
\end{equation}
When the chain is conditioned upon the number $n$ of sampled individuals, it remains a Markov chain $((R_i,S_i);1\le i\le n)$, but the transition probability becomes $p(x;dy\,dz)/(1-k(x))$, which now integrates to 1.
\end{thm}
\amaury{The formula for the transition probability is a direct consequence, by elementary calculus, of Proposition \ref{prop : prelim} and Lemmas \ref{lem1} and \ref{lem2}. The remaining statements are proved in} \helen{the Appendix.} 

In the rest of this section, \emph{we assume that $V_2$ has a density}, say $g_2$, in the sense that $P(V_2\in du)=g_2(u)\, du$, so that $C_1^{(q)}$ is differentiable with derivative 
\begin{equation}
\label{C'}
C_1^{(q)\prime}(z) = q\,g_2(z)+q\int_0^z W_1^{(q)\prime}(z-u)\, g_2(u)\, du,
\end{equation}
\amaury{where the first term comes from differentiating the integral as a function of its upper bound and the second one comes from differentiating the function of $z$ inside the integral}.
The first consequence is that $S_1$ has a density, say $g$, given by
\begin{equation}
\label{g}
g(z)=b^{-1}\,\left(C_1^{(q)\prime}(z)-\frac{C_1^{(q)}(z)C_1^{(q)}(t)}{U_1^{(q)}(t)}\right).
\end{equation}
The second consequence is that the transition probability has density, say $f$,
$$
\amaury{P_x(R\in dy, S\in dz)=} p(x;dy\,dz)=: f(x;y,z)\,dy\, dz,
$$
where, by differentiating the expression given in Theorem \ref{thm} for $P_x(R> y, S \in dz)/dz$ \amaury{with respect to $y$ and recalling that $U_1^{(q)\prime}=C_1^{(q)}$}, we get
\begin{equation}
\label{f}
f(x;y,z)
= \frac{U_1^{(q)}(t-x)}{U_1^{(q)}(t-y)}\,\left[C_1^{(q)\prime} (z-y) - C_1^{(q)}(z-y)\,\frac{C_1^{(q)}(t-y)}{U_1^{(q)}(t-y)}\right] .
\end{equation}

Then we can directly write down the likelihood of a given oriented tree as follows.

\begin{cor}
\label{cor}
For any given oriented tree $\mathcal{T}$ with coalescence times $(y_i)_{2\le i\le n}$ and sampling times $(z_i)_{1\le i\le n}$, where tips are labeled from left to right, the likelihood $\mathcal{L}_S (\mathcal{T})$ of this tree under the general epidemiological model observed at time $t$, conditional on at least one sampled individual, is
$$
\mathcal{L}_S(\mathcal{T}) =  \frac{g(z_1)\,k(z_n)}{p} \,\prod_{i=2}^n f(z_{i-1}; y_i, z_i),
$$
where $k$ and $p$ are given by \eqref{k} and \eqref{p} in Theorem \ref{thm}, and $g$ and $f$ are given respectively by \eqref{g} and \eqref{f}. 

Alternatively, we can condition on the number $n$ of sampled individuals.  Applying the remark in Theorem \ref{thm}, we obtain the conditional likelihood $\mathcal{L}_n(\mathcal{T})$
$$
\mathcal{L}_n(\mathcal{T}) = \frac{g(z_1)\,k(z_n)}{p}\,\prod_{i=2}^n \frac{f(z_{i-1}; y_i, z_i)}{1-k(z_{i-1})},
$$

\end{cor}

\section{Worked examples}

For illustration, we now describe two specific cases of the general model, meant as simplistic descriptions  of influenza and HIV epidemics, respectively.  We apply our mathematical results to these cases, by specifically deriving the expressions required for the likelihood.

\subsection{Influenza}
In the case of influenza, we assume that, after a random amount of time, an infective either recovers without sampling, with a certain probability $c_1$ which does not depend on the time elapsed, or with probability $c_2$, recovers with sampling, which typically happens for the severe cases in the hospital.
Thus we assume that $V_1$ and $V_2$ are equal in distribution.
The following statement is a straightforward consequence of Lemma \ref{lem:useful} and is needed for practical applications of Theorem \ref{thm}. In the case when $V_1=V_2$ has a density, it is recommended to use Equations \eqref{C'}, \eqref{g}, \eqref{f} for such practical applications.

\begin{prop}
In the influenza model, we have
$$
C_1^{(q)}(z) = (c_2/c_1)(bW_1^{(q)}(z)-W_1^{(q)\prime}(z)),
$$
so that
$$
U_1^{(q)}(z) = 1+(c_2/c_1) \left( 1+b\int_0^zW_1^{(q)}(s)\, ds-W_1^{(q)}(z) \right).
$$
\end{prop}

\subsection{HIV}
\label{HIV}
In the case of HIV, each individual has a ``natural'' infectious lifetime, denoted $V$, having an arbitrary distribution.  This lifetime would apply if there were no intervention. However, individuals are additionally sampled after some independent exponential duration, say $\mathbf{e'}$, with rate parameter $\mu$.  Individuals are removed from the infectious class upon sampling, due e.g.~to successful treatment or behavior change concomitant with the intervention.  We note that a model with constant rates of both ``natural death'' and sampling, as in \cite{Stadler2012MBE-R0}, is a special case of this model; see the computations for the Markovian case at the end of this section.

Setting $V^\mu:=\min(V, \mathbf{e}')$, where $V$ and $\mathbf{e'}$ are assumed independent, we have the probability of sampling:
$$ c_2=P(V^\mu =\mathbf{e'}) = P(\mathbf{e'}<V) = 1-E(e^{-\mu V})
$$
which we can rewrite as:
\begin{equation}
c_2=P(V>\mathbf{e'}) = \int_{(0,\infty]}\mu\,e^{-\mu r}P(V>r) dr= 1-\int_{(0,\infty]} e^{-\mu r}P(V\in dr)= 1-c_1.
\label{eq : c2re}
\end{equation}
Furthermore,
\begin{equation}
\label{eqn : V1V2}
P(V_1\in dr) = c_1^{-1}\, e^{-\mu r}P(V\in dr)\quad\mbox{ and }\quad P(V_2\in dr) = c_2^{-1}\,\mu\,e^{-\mu r} P(V>r) dr.
\end{equation} 
Taking $\psi(\lambda):=\lambda - b\intgen (1-e^{-\lambda r}) P(V\in dr)$ and manipulating Equation \eqref{eq : c2re} yields 
$$c_2=\frac{\mu-\psi(\mu)}{b},$$
while,
$$
\psi_1(\lambda) := \lambda -bc_1 \intgen (1-e^{-\lambda r}) P(V_1\in dr)=\lambda -bc_1 +b\intgen e^{-(\lambda+\mu) r}P(V\in dr)= \psi(\lbd+\mu) - \psi(\mu).
$$
Now since $\psi_1(\lbd)=\psi(\lbd+\mu) - \psi(\mu)$ and $q=b c_2 = \mu-\psi(\mu)$, notice that $\psi_1(\lambda)-q=\psi(\lambda+\mu)-\mu$.

The following statement is  needed for practical applications of Theorem \ref{thm}. The proof is found in the appendix.
\begin{prop} \label{PropHIV}
In the HIV model, we have
$$
C_1^{(q)}(z) = \mu\int_0^z e^{-\mu x} W_1^{(q)\prime}(z-x)\, dx =  \mu\left(W_1^{(q)}(z)-1-\int_0^z\mu\,e^{-\mu x} \left( W_1^{(q)}(z-x)-1\right)\, dx\right),
$$
$$
U_1^{(q)}(z) = 1+\mu \int_0^z dx\,e^{-\mu x}\left(W_1^{(q)}(z-x) - 1\right) = W_1^{(q)}(z) - \mu^{-1}C_1^{(q)}(z),
$$
and the initial distribution of $S_1$ is given by:
$$
P(S_1\in dz) = \frac{\mu}{b}\left(W_1^{(q)\prime}(z) - \frac{C_1^{(q)}(z)W_1^{(q)}(t)}{U_1^{(q)}(t)} \right)\, dz.
$$
\end{prop}

We make further computations in the Markovian case, that is, when the ``natural'' lifetime of individuals ends at constant rate $d$. Then $\pi(dr)=bde^{-dr}\,dr$ 
and
$$
\psi(\lbd+\mu) - \mu = \frac{\amaury{Q}(\lbd)}{\lbd+\mu+d} ,
$$
where $\amaury{Q}(\lbd)=\lbd^2+\lbd(\mu+d-b) -b\mu$. Then the polynomial \helen{$Q$} has  two distinct real roots
$$
-\alpha_1=\left(b-d-\mu-\sqrt{(\mu+d -b)^2+4b\mu}\right)/2\quad\mbox{ and }\quad \alpha_2=\left(b-d-\mu+\sqrt{(\mu+d -b)^2+4b\mu}\right)/2,
$$
where $\alpha_1$ and $\alpha_2$ are both positive. Using $\alpha_2-\alpha_1=b-d-\mu$, we get
$$
\frac{1}{\psi(\lbd+\mu)-\mu} = \frac{1}{\alpha_1+\alpha_2}\left[\frac{\alpha_2-b}{\lbd+\alpha_1}+ \frac{\alpha_1+b}{\lbd-\alpha_2}\right] ,
$$
so that
$$
W_1^{(q)}(x) = \frac{\alpha_2-b}{\alpha_1+\alpha_2}\,e^{-\alpha_1 x}+ \frac{\alpha_1+b}{\alpha_1+\alpha_2}\,e^{\alpha_2 x}\qquad x\ge 0.
$$
We demonstrate in the Appendix that applying Theorem \ref{thm} leads to the same expression for the likelihood as derived previously using methods particular to this Markovian case \cite{Stadler2012MBE-R0}.

\section{Discussion}
We introduced a stochastic population dynamics model giving rise to phylogenetic trees with sequentially sampled tips. The lifetime of the individuals within the population may follow an arbitrary distribution, while the production of ``daughter'' individuals occurs with a constant rate.
We showed that the two-dimensional coalescent point process formed by pairs of coalescence and sampling times in the left-to-right order satisfies the Markov property.
We characterized the law of this Markov chain, providing a framework to calculate the likelihood of a phylogenetic tree, as displayed in Theorem \ref{thm} and especially in Corollary \ref{cor}.

Evaluating the likelihood of a phylogenetic tree requires the numerical evaluation of the function $W_1^{(q)}$. 
This evaluation can be performed either by solving the inverse Laplace transform in \eqref{eqn : LT scale} or the integro-differential equation in Lemma \ref{lem:useful}.
We leave the numerical challenges for a future study. However, for the special case of exponentially distributed lifetimes, analytic solutions for the inverse Laplace transform and thus also for the likelihood of the sampled tree are available \cite{Stadler2010JTB,Stadler2012MBE-R0}. A special section is dedicated to this case in the Appendix.

\tanja{We envision to use the model on epidemiological data in the following way.  Pathogen genetic sequencing data from different hosts is used to reconstruct the genealogical relationship of the data, i.e. the phylogenetic tree. This phylogenetic tree is  treated as a proxy for the transmission tree (i.e. branching events are transmission events). We do not deal with this reconstruction and assume for our method that the reconstructed tree is provided. We then assume that the model introduced in this paper gave rise to the transmission tree, and want to fit the model to the tree using the  likelihood function. There are two ways to do the fitting.
First, the likelihood of the tree can be used for determining maximum likelihood parameter estimates for a given sampled phylogenetic tree, by maximizing the probability of the sampled tree over the parameters.
Second, the likelihood together with prior distributions on the model parameters can be used in a Bayesian framework to obtain the posterior distribution of parameters given a sampled tree. }

We stress that real data (i.e. sequences, sampling times and/or the associated phylogenetic tree) do not come with the information on the orientation of the tree. However, different orientations lead to different likelihoods, since different orientations can give rise to different precise pairings of successive coalescence and sampling times ($R$ and $S$).  An additional computational challenge is thus to sum the likelihood over all valid $(R,S)$ pairings.

The second useful application of our framework is concerned with the simulation of phylogenetic trees. If simulating the model forward in time, one must simulate many non-sampled individuals, and thus it takes much computational time to obtain the required number of samples. However, using the Markov chain property of our coalescent and sampling time pairs, we can sample once from the distribution for the starting point and $n-1$ times from the distribution specifying the Markov chain in order to obtain a tree on $n$ tips.

So far we had to assume a constant birth rate. Generalizing the results to time-dependent birth rates, \amauryy{as well as death/sampling rates}, should be conceptually straightforward: the ball in the contour process is simply rolling back towards 0 with a varying speed.  However, generalizing to age-dependent birth rates, i.e.~an arbitrary distribution of time until birth of a new individual, will most likely be unachievable with the current framework, as we can no longer let the ball roll back without knowing the age of the individual it represents.

We conclude by emphasizing that our analyses were performed with an epidemiological application in mind; however, any implementation may also be useful for analyzing phylogenetic trees with sequentially sampled tips arising in different applications, such as species phylogenies with fossil tips.

\paragraph{Acknowledgments.} AL was financially supported by grant MANEGE `Mod\`eles Al\'eatoires en \'Ecologie, G\'en\'etique et \'Evolution' 09-BLAN-0215 of ANR (French national research agency). AL also thanks the {\em Center for Interdisciplinary Research in Biology} (Collège de France) for funding. HA received support from ETH Z\"urich. TS thanks the Swiss National Science foundation for funding (SNF grant \#PZ00P3 136820).



\appendix

\section{Proofs}

\subsection{Proof of Lemma \ref{lem:useful}}

By an integration by parts, the Laplace transform (as a function of $\lambda>\phi(q)$) of the non-negative function $z\mapsto W^{(q)\prime}(z)+\int_0^z W^{(q)}(z-x) \,\pi(dx)$ equals
$$
[e^{-\lambda z} W^{(q)}(z)]_0^\infty + \frac{\lambda}{\psi(\lambda)-q}+ \frac{\int_0^\infty\pi(dx)\,e^{-\lambda x}}{\psi(\lambda)-q}=-1+ \frac{\lambda}{\psi(\lambda)-q}+\frac{\psi(\lambda)-\lambda +b}{\psi(\lambda)-q} =\frac{q +b}{\psi(\lambda)-q}, 
$$
\amaury{where we used successively the facts  that the Laplace transform of $W^{(q)}$ is $1/(\psi(\lambda)-q)$, that the Laplace transform of a convolution product is the product of Laplace transforms, and that $W^{(q)}(0)=1$.} Now the right-hand side is also the Laplace transform of the non-negative function $z\mapsto(q+b)W^{(q)} (z)$. \hfill $\Box$\\

\subsection{Proof of Lemma \ref{lem1}}
Set
$$
H^{(q)}(x,t;dz):=P_x(\If_{\mathbf{e}} > 0, \Sp_{\mathbf{e}}\le t, X_{\mathbf{e}}+V_2\in dz).
$$
By  \eqref{eqn : q-resolvent-bis}, defining $u^{q}_t$ the $q$-resolvent of the process $X$ killed upon exiting $(0,t]$, we get
$$
H^{(q)}(x,t;dz)=q\, \int_0^z u^{q}_t(x,dr)\,P(V_2 \in dz -r)
$$
so by Equations \eqref{eqn : q-resolvent} and \eqref{eqn : dfn C},
$$
H^{(q)}(x,t;dz)/dz=\frac{W_1^{(q)}(t-x)}{W_1^{(q)}(t)}\, C_1^{(q)}(z) - \indic{z\ge x}\,C_1^{(q)}(z-x),
$$

In conclusion,
\begin{equation}
\label{eqn:resolvent+V2}
P_x(\If_{\mathbf{e}} > 0, \Sp_{\mathbf{e}}\le t, X_{\mathbf{e}}+V_2\in dz)=\left(\frac{W_1^{(q)}(t-x)}{W_1^{(q)}(t)}\,C_1^{(q)}(z) - \indic{z\ge x}\, C_1^{(q)}(z-x)\right) \, dz.
\end{equation}
Invariance by translation yields the result.

\subsection{Proof of Lemma \ref{lem2}}

Integrating over $z$ the equality in the previous lemma and applying Equation \eqref{eqn : dfn U} yields
\begin{equation}
\label{eqn : L1re}
P_x(\If_{\mathbf{e}} > y, \Sp_{\mathbf{e}}\le t, X_{\mathbf{e}}+V_2\le z)=
\frac{W_1^{(q)}(t-x)}{W_1^{(q)}(t-y)}(U_1^{(q)}(z-y)-1)-\indic{z\ge x}(U_1^{(q)}(z-x)-1).
\end{equation}
Noting that $\tau \equiv \mathbf{e}$ on $C'$, we deduce
\begin{eqnarray*}
P_x(\If_\tau >y, C')
	&=& P_x(\If_{\mathbf{e}}>y, \Sp_{\mathbf{e}}\le t, X_{\mathbf{e}}+V_2>t)\\
	&=& P_x(\If_{\mathbf{e}}>y, \Sp_{\mathbf{e}}\le t) - P_x(\If_{\mathbf{e}}>y, \Sp_{\mathbf{e}}\le t, X_{\mathbf{e}}+V_2\le t)\\
	&=& P_x\left( \mathbf{e}< T_{y}\wedge T_{(t,+\infty)}\right) - \frac{W_1^{(q)}(t-x)}{W_1^{(q)}(t-y)}(U_1^{(q)}(t-y)-1)+(U_1^{(q)}(t-x)-1).
\end{eqnarray*}
where the last equality follows by applying \eqref{eqn : L1re} with $z=t$.
On the other hand,
\begin{eqnarray*}
P_x(\If_\tau >y, C)
&=& P_x\left(  T_{(t,+\infty)}<T_{y}\wedge \mathbf{e} \right)\\
&=& 1-P_x\left( \mathbf{e}< T_{y}\wedge T_{(t,+\infty)}\right)-P_x\left(  T_{y}<T_{(t,+\infty)}\wedge \mathbf{e} \right)\\
&=& 1-P_x\left( \mathbf{e}< T_{y}\wedge T_{(t,+\infty)}\right)-\frac{W_1^{(q)}(t-x)}{W_1^{(q)}(t-y)},
\end{eqnarray*}
where the last equality is due to \eqref{eqn : q-two-sided}.

Since $C$ and $C'$ are mutually exclusive, we can sum the last two sets of equations to obtain
\begin{eqnarray*}
P_x(\If_\tau >y, C\cup C')
&=&1-\frac{W_1^{(q)}(t-x)}{W_1^{(q)}(t-y)}- \frac{W_1^{(q)}(t-x)}{W_1^{(q)}(t-y)}(U_1^{(q)}(t-y)-1)+U_1^{(q)}(t-x)-1\\
&=&-\frac{W_1^{(q)}(t-x)}{W_1^{(q)}(t-y)}\,U_1^{(q)}(t-y)+U_1^{(q)}(t-x),
\end{eqnarray*}
which was the announced result.

\subsection{Proof of Theorem \ref{thm}}
\label{subsec:A3}
Recall that the formula for the transition probability is a direct consequence of Proposition \ref{prop : prelim} and Lemmas \ref{lem1} and \ref{lem2}.

The computation of the killing probability can be obtained by two methods. The first method uses the formula in Proposition \ref{prop : prelim}. Taking $y=0$ in Lemma \ref{lem2}, we get 
$$
P_x(C\cup C')
= U_1^{(q)}(t-x)  - \frac{W_1^{(q)}(t-x)}{W_1^{(q)}(t)}U_1^{(q)}(t) .
$$
Also $P_x(A\cup B)+ P_x(C\cup C')=1$ and by \eqref{eqn : q-resolvent-bis},
$$
P_x(A) = \frac{W_1^{(q)}(t-x)}{W_1^{(q)}(t)},
$$
which suffices to terminate the computation. The second method uses the fact that $1-k(x)$ is the total mass of the measure $p(x;\cdot)$. Taking $y=0$ in the transition probability, one gets
\begin{equation}
\label{eqn : law S}
P_x(S \in dz) = \left(C_1^{(q)}(z)\,\frac{U_1^{(q)}(t-x)}{U_1^{(q)}(t)}-\indic{z\ge x}C_1^{(q)}(z-x)\right)\, dz .
\end{equation}
The present alternative proof ends integrating the last density over $[0,t]$ and using \eqref{eqn : dfn U}.

As a last step, we express the distribution of $S_1$. To compute the law of $S_1$, observe that either the progenitor of the genealogy is sampled before $t$, or otherwise, conditional on the lifetime $x$ of this progenitor, $S_1$ is distributed according to $P_x(S\in\cdot)$. This can be written as follows, integrating over the different possible values of $x$, greater than $t$ (in which case reflection occurs) or smaller than $t$:
$$
P(S_1\in dz) = c_2\, P(V_2 \in dz)+(c_1 P(V_1\ge t)+c_2 P(V_2\ge t))P_t (S\in dz)+\int_{(0,t)}c_1 \,P(V_1\in dr)\,P_r (S\in dz).
$$
From \eqref{eqn : law S}, we get, after some algebra,
$$
P(S_1\in dz) = c_2P(V_2\in dz) + \frac{C_1^{(q)}(z)}{U_1^{(q)}(t)} \left( c_1 P(V_1\ge t) + c_2P(V_2\ge t) \right)+b^{-1}A^{(q)}(t))\, dz - b^{-1}B^{(q)}(z) \, dz,
$$
where
\begin{multline*}
A^{(q)}(t) := bc_1\int_{(0,t)}P(V_1\in dr) U_1^{(q)}(t-r) \\= bc_1P(V_1<t) + bc_1q\int_0^t P(V_1\in dr)\int_0^{t-r}W_1^{(q)}(t-r-u)P(V_2\le u)\, du,
\end{multline*}
and
$$
B^{(q)}(z) := bc_1 \int_{(0,z)}P(V_1\in dr) C_1^{(q)}(z-r) = bc_1q\int_{(0,z)}P(V_1\in dr) \int_0^{z-r}W_1^{(q)}(z-r-u)P(V_2\in du).
$$
Using the commutativity of the convolution product and Lemma \ref{lem:useful}, and recalling that $q= bc_2 = b(1-c_1)$, we get
\begin{eqnarray*}
A^{(q)}(t) &=& bc_1P(V_1<t) + q\int_0^t du\, P(V_2< u)\int_0^{t-u}W_1^{(q)}(t-r-u)bc_1P(V_1\in dr)\\
 &=&bc_1P(V_1<t) + q\int_0^t du\, P(V_2< u)(bW_1^{(q)}(t-u)- W_1^{(q)\prime}(t-u))\\
 &=&bc_1P(V_1<t) + b(U_1^{(q)}(t)-1)- q\int_0^t du\, P(V_2< u)W_1^{(q)\prime}(t-u)\\
 &=&bc_1P(V_1<t) + b(U_1^{(q)}(t)-1)+ bc_2 P(V_2<t)-q\int_0^t  P(V_2\in du)W_1^{(q)}(t-u)\\ 
 &=&b(c_1P(V_1<t) + c_2 P(V_2<t)) +b(U_1^{(q)}(t)-1)-C_1^{(q)}(t).
\end{eqnarray*}
Similarly,
\begin{eqnarray*}
B^{(q)}(z) &=&  q\int_0^z du\, P(V_2\in du)\int_0^{z-u}W_1^{(q)}(z-r-u)bc_1P(V_1\in dr)\\
 &=&q\int_0^z  P(V_2\in du)(bW_1^{(q)}(z-u)- W_1^{(q)\prime}(z-u))\\
 &=& bC_1^{(q)}(z)- q\int_0^z  P(V_2\in du) W_1^{(q)\prime}(z-u).
\end{eqnarray*}
Substituting the final expressions for $A^{(q)}$ and $B^{(q)}$ into the previous expression for $P(S_1\in dz)$ finally yields:
$$
P(S_1\in dz) = c_2P(V_2\in dz) + \left(c_2 \int_0^z  P(V_2\in du)\, W_1^{(q)\prime}(z-u)-\frac{C_1^{(q)}(z)C_1^{(q)}(t)}{bU_1^{(q)}(t)}\right)\, dz,
$$
which was to be proved. 

Finally, the probability $p$ that at least one individual is sampled before time $t$ is given by
\begin{eqnarray*}
p&=&\int_0^t P(S_1\in dz)\\ 
	&=& c_2P(V_2\le t) + c_2 \int_0^t dz \int_0^z  P(V_2\in du)\, W_1^{(q)\prime}(z-u)-\frac{C_1^{(q)}(t)}{bU_1^{(q)}(t)}\int_0^t C_1^{(q)}(z)\, dz\\
	&=& c_2P(V_2\le t) + c_2 \int_0^t P(V_2\in du)\, \left(   W_1^{(q)}(t-u)-1\right)-\frac{C_1^{(q)}(t)}{bU_1^{(q)}(t)} \left(U_1^{(q)}(t)-1\right)\\	
	&=& c_2P(V_2\le t) + c_2 \int_0^t P(V_2\in du)\,   W_1^{(q)}(t-u)- c_2 P(V_2\le t)-b^{-1}C_1^{(q)}(t) + \frac{C_1^{(q)}(t)}{bU_1^{(q)}(t)} \\
	&=& \frac{C_1^{(q)}(t)}{bU_1^{(q)}(t)} ,
\end{eqnarray*}
which is the announced result.\hfill $\Box$

\subsection{Proof of Proposition \ref{PropHIV}}
First, using the convolution rule for Laplace transforms and then Equation \eqref{eqn : V1V2}, the Laplace transform (as a function of $\lambda$) of $C_1^{(q)}$ is
\begin{eqnarray*}
\frac{qE(e^{-\lambda V_2})}{\psi_1(\lambda) -q} 
	&=& \frac{\intgen b\mu e^{-\mu r} P(V>r) e^{-\lambda r}\, dr}{\psi_1(\lambda) -q} \\
	&=& \frac{b\mu}{\lambda + \mu}\,\frac{1-E(e^{-(\lambda+\mu)V})}{\psi_1(\lambda) -q}\\
	&=& \frac{\mu}{\lambda + \mu}\,\frac{\lambda+\mu - \psi(\lambda+\mu)}{\psi_1(\lambda) -q}\\
	&=& \frac{\mu}{\lambda + \mu}\,\left(-1+\frac{\lambda}{\psi_1(\lambda) -q}\right).	
\end{eqnarray*}
Now since the first factor in the final product is the Laplace transform of the exponential density with parameter $\mu$ and the second factor is the Laplace transform of $W_1^{(q)\prime}$, we get (by the convolution rule) the first proposed expression for $C_1^{(q)}$.  The second one follows by an integration by parts.  By substituting the first expression for $C_1^{(q)}$ into Equation \eqref{eqn : dfn U}, one obtains the first expression proposed for $U_1^{(q)}(z)$.  The second follows by rearranging terms in the second expression for $C_1^{(q)}$.

Let us now compute the initial distribution of $S_1$. To this end, we compute an expression for $I(z):= \mu^{-1}c_2 \int_0^z P(V_2\in du)\, W_1^{(q)\prime} (z-u)$. Applying Equation \eqref{eqn : V1V2} (the laws of $V_1$ and $V_2$), we get
\begin{eqnarray*}
 I(z) &=& \int_0^z e^{-\mu u} P(V>u) W_1^{(q)\prime}(z-u)\, du\\
	&=& [-W_1^{(q)}(z-u)  e^{-\mu u} P(V>u)]_0^z -\int_0^z W_1^{(q)}(z-u)(c_2 P(V_2\in du) + c_1 P(V_1\in du))\\
	&=& -\mu^{-1}c_2 P(V_2\in dz)/dz+W_1^{(q)}(z) -b^{-1} C_1^{(q)}(z) - b^{-1}(bW_1^{(q)}(z) -W_1^{(q)\prime}(z)), 
\end{eqnarray*}
where the second equality is an integration by parts and the last one is due to Lemma \ref{lem:useful} and Equation \eqref{eqn : dfn C}. Then we get
$$
\mu I(z) = c_2 \int_0^z P(V_2\in du)\, W_1^{(q)\prime} (z-u) =-c_2 P(V_2\in dz)/dz +\mu b^{-1} (-C_1^{(q)}(z) +W_1^{(q)\prime}(z)).
$$
Using the general expression for the initial distribution of $S_1$ in Theorem \ref{thm}, we get
\begin{eqnarray*}
P(S_1\in dz)/dz &=&  \mu b^{-1}(- C_1^{(q)}(z)+W_1^{(q)\prime}(z)) -\frac{C_1^{(q)}(z)C_1^{(q)}(t)}{bU_1^{(q)}(t)}\\
	&=& \mu b^{-1}W_1^{(q)\prime}(z) - \mu b^{-1}  C_1^{(q)}(z)\frac{U_1^{(q)}(t)+\mu^{-1}C_1^{(q)}(t)}{U_1^{(q)}(t)}\\
	&=& \frac{\mu}{b}\left(W_1^{(q)\prime}(z) - \frac{C_1^{(q)}(z)W_1^{(q)}(t)}{U_1^{(q)}(t)} \right),
\end{eqnarray*}
which ends the proof.\hfill $\Box$\\

\subsection{Likelihood in the Markovian case}

In the Markovian case, individuals die at constant rate $d$ and are sampled at constant rate $\mu$.  In this competing-exponentials case, we have $c_2 = \frac{\mu}{\mu+d}$ and $P(V_2 \in dr)=(\mu+d) e^{-(\mu+d)r} dr$.  The scale function $W_1^{(q)}$ was already presented in Section \ref{HIV}, and we now compute the remaining functions required for the expression of the likelihood.  To obtain simple expressions, we note the following useful relationships, where $\alpha_1$ and $\alpha_2$ are defined in Section \ref{HIV}.
\begin{align*}
\alpha_1 \alpha_2 &= b \mu \\
(\alpha_1-\mu-d)(\alpha_2+\mu+d) &= -bd \\
(\alpha_1+b)(\alpha_1-\mu-d) &= -bd \\
(\alpha_2-b)(\alpha_2+\mu+d) &= -bd \\
\alpha_1 (\alpha_2 - b)(\alpha_2 + \mu) &= -b d \mu \\
\alpha_2 (\alpha_1 + b) (\alpha_1 - \mu) &= b d \mu
\end{align*}
Then, using the definitions of $C_1^{(q)}$ and $U_1^{(q)}$ in Equations \eqref{eqn : dfn C} and \eqref{eqn : dfn U}, and simplifying, we obtain in summary:
\begin{align*}
W_1^{(q)}(x) &= \frac{\alpha_2-b}{\alpha_1+\alpha_2} e^{-\alpha_1 x} + \frac{\alpha_1+b}{\alpha_1+\alpha_2} e^{\alpha_2 x} \\
W_1^{(q)\prime}(x) &= \frac{b}{\alpha_1+\alpha_2} \left( (\alpha_1-\mu) e^{-\alpha_1 x} + (\alpha_2+\mu) e^{\alpha_2 x} \right) \\
C_1^{(q)}(x) &= \frac{b \mu}{\alpha_1+\alpha_2} \left( e^{\alpha_2 x} - e^{-\alpha_1 x} \right) \\
C_1^{(q)\prime}(x) &=\frac{b \mu}{\alpha_1+\alpha_2} \left( \alpha_1 e^{-\alpha_1 x} + \alpha_2 e^{\alpha_2 x} \right) \\
U_1^{(q)}(x) &= \frac{\alpha_2 e^{-\alpha_1 x} + \alpha_1 e^{\alpha_2 x}}{\alpha_1+\alpha_2}
\end{align*}
We can now proceed to calculate the factors involved in the likelihood (Corollary \ref{cor}).  Substituting the required functions and simplifying, we have
\begin{align*}
g(z_1) &= \frac{\mu e^{\alpha_2 z_1} \left( \alpha_2+\alpha_1 e^{(\alpha_1+\alpha_2) (t-z_1)} \right)}{\alpha_2+\alpha_1 e^{(\alpha_1+\alpha_2) t}} \\
k(z_n) &= \frac{e^{\alpha_1 z_n} \left( \alpha_2+\alpha_1 e^{(\alpha_1+\alpha_2)(t-z_n)} \right)}{\alpha_2+\alpha_1 e^{(\alpha_1+\alpha_2) t}} \\
p &= \frac{\mu (e^{(\alpha_1+\alpha_2)t}-1)}{\alpha_2 + \alpha_1 e^{(\alpha_1+\alpha_2)t}} \\
f(z_{i-1};y_i,z_i) &= \frac{b \mu e^{\alpha_1 (z_{i-1}-y_i)} e^{\alpha_2 (z_i-y_i)} \left( \alpha_2 + \alpha_1 e^{(\alpha_1+\alpha_2)(t-z_{i-1})} \right) \left( \alpha_2 + \alpha_1 e^{(\alpha_1+\alpha_2)(t-z_i)} \right)}{\left( \alpha_2+\alpha_1 e^{(\alpha_1+\alpha_2) (t-y_i)} \right) ^2} 
\end{align*}

For direct comparison to the likelihood derived previously for the Markovian case \cite{Stadler2012MBE-R0}, we consider the likelihood given the time of observation ($t$) but not conditioned on sampling, which we denote $\mathcal{L}(\mathcal{T})$.  Substituting the above factors and simplifying, we have
\begin{align}
\mathcal{L}(\mathcal{T}) &= g(z_1) k(z_n) \prod_{i=2}^n f(z_{i-1};y_i,z_i) \nonumber \\
&= b^{n-1} \mu^n \frac{1}{e^{-(\alpha_1+\alpha_2) t} \left( \alpha_2 + \alpha_1 e^{(\alpha_1+\alpha_2) t} \right)^2} \frac{\prod_{i=1}^n e^{-(\alpha_1+\alpha_2) (t-z_i)} \left( \alpha_2 + \alpha_1 e^{(\alpha_1+\alpha_2)(t-z_i)} \right)^2}{\prod_{i=2}^n e^{-(\alpha_1+\alpha_2) (t-y_i)} \left( \alpha_2 + \alpha_1 e^{(\alpha_1+\alpha_2)(t-y_i)} \right)^2}
\label{eqn : LlhdA}
\end{align}
On the other hand, the likelihood was previously derived \cite{Stadler2012MBE-R0} as the following, adjusted to match present notation:
\begin{equation}
\label{eqn : LlhdT}
\mathcal{L}(\mathcal{T}) = b^{n-1} \mu^n \frac{1}{q(t)} \frac{\prod_{i=1}^n q(t-z_i)}{\prod_{i=2}^n q(t-y_i)}
\end{equation}
with the definitions
\begin{align*}
q(x) = 2(1-\gamma_2^2) + \mathrm{e}^{-\gamma_1 x} (1-\gamma_2)^2 + \mathrm{e}^{\gamma_1 x} (1+\gamma_2)^2 ,\\
\gamma_1 = \sqrt{(b-d-\mu)^2 + 4 b \mu}, \quad \gamma_2 = -\frac{b-d-\mu}{\gamma_1}
\end{align*}
Note that $\gamma_1 = \alpha_1+\alpha_2$ and $\gamma_2 = \frac{\alpha_1-\alpha_2}{\alpha_1+\alpha_2}$.  We can thus rewrite,
$$ q(x) = \frac{4 e^{-(\alpha_1 +\alpha_2) x}}{(\alpha_1 +\alpha_2)^2} \left( \alpha_2 + \alpha_1 e^{(\alpha_1 +\alpha_2) x} \right)^2
$$
Cancelling the constant factors in $q(\cdot)$, it immediately follows that Equations \eqref{eqn : LlhdA} and \eqref{eqn : LlhdT} precisely agree.

\clearpage

\clearpage

\end{document}